\documentclass[runningheads]{llncs}
\usepackage{graphicx}
\usepackage{amssymb}
\usepackage{amsmath}
\begin{document}
\title{Privacy Preserving Multi-Server $k$-means Computation over Horizontally Partitioned Data}
\titlerunning{Privacy Preserving Multi-Server $k$-means Computation}
%
\author{Riddhi Ghosal\inst{1} \and
Sanjit Chatterjee\inst{2}}
\authorrunning{R. Ghosal and S. Chatterjee}
%
\institute{Indian Statistical Institute \\
\email{postboxriddhi@gmail.com}\\ \and
Department of Computer Science and Automation, Indian Institute of Science\\
\email{sanjit@iisc.ac.in}}
\maketitle              
\begin{abstract}
The $k$-means clustering is one of the most popular clustering algorithms in data mining. Recently a lot of research has been concentrated on the algorithm when the data-set is divided into multiple parties or when the data-set is too large to be handled by the data owner. In the latter case, usually some servers are hired to perform the task of clustering. The data set is divided by the data owner among the servers who together compute the $k$-means and return the cluster labels to the owner. The major challenge in this method is to prevent the servers from gaining substantial information about the actual data of the owner. Several algorithms have been designed in the past that provide cryptographic solutions to perform privacy preserving $k$-means. We propose a new method to perform $k$-means over a large set of data using multiple servers. Our technique avoids heavy cryptographic computations and instead we use a simple randomization technique to preserve the privacy of the data. The $k$-means computed has essentially the same efficiency and accuracy as the $k$-means computed over the original data-set without any randomization. We argue that our algorithm is secure against honest-but-curious and and non-colluding adversary.
\end{abstract}
\section{Introduction}
\label{S:1}
The $k$-means clustering is one of the most widely used techniques in data mining 
\cite{hartigan1979algorithm,kanungo2002efficient,alsabti1997efficient,likas2003global,jain2010data}. The $k$-means clustering algorithm is used to find groups which have not been explicitly labeled in the data. This can be used to confirm business assumptions about what types of groups exist or to identify unknown groups in complex data sets.  It has been successfully used in various domains including market segmentation, computer vision, geostatistics, astronomy and agriculture 
\cite{oyelade2010application,tellaeche2007vision,celik2009unsupervised}. $k$-means clustering is rather easy to implement and apply even on large data sets, particularly when using heuristics such as Lloyd's algorithm. However, sometimes  the data-set contains private information that cannot be made available to the party who is computing the $k$-means for a user 
\cite{Cranor:1999:IP:293411.293440,agrawal2000privacy}. There are times when the data is huge and the data owner does not have the computational capability to do clustering on his/her own. In our work we deal with this particular case. Another scenario
may be a few independent parties contain parts of data on whom clustering has to be performed as a whole 
\cite{Bunn:2007:STK:1315245.1315306,Vaidya:2003:PKM:956750.956776,jagannathan2006new}. 

The privacy and secrecy considerations can prohibit the parties from sharing their data with each other. The solution should not just provide the required privacy assurance  but should also minimize the additional overheads in terms of communication and computation costs required to introduce privacy. Solutions proposed in the works such as 
\cite{Bunn:2007:STK:1315245.1315306,jagannathan2006new,Vaidya:2003:PKM:956750.956776,Doganay:2008:DPP:1379287.1379291}
compute $k$-means by making the participating parties compute common functions, without having to actually reveal their individual data to any other party. Such algorithms face a lot of challenges because it is not very easy to reach an optimal point that will provide a perfect balance for security, accuracy and efficiency.

One of the most common approaches to solve this issue is using data perturbation to preserve the privacy of the data. 
Some of the common techniques are using additive noise \cite{kargupta2003privacy},  multiplicative noise \cite{10026864614}, 
geometric perturbation, or rotational perturbation \cite{Chen2005PrivacyPD}, all of which have the ``Distance Preservation Property". 
Some works use secure multiparty computation \cite{Samet2007PrivacyPK}, and 
homomorphic encryptions \cite{Beye2011EfficientPP,Jagannathan:2005:PDK:1081870.1081942} to safeguard the data. But these schemes are generally computationally costly and reduce the performance of the clustering algorithm significantly. The latter approaches provide more protection to the data than the former at the cost of efficiency and sometimes their application becomes practically infeasible.

The setup we consider in this work is somewhat similar to \cite{Upmanyu:2010:EPP:2163922.2163944} in which \textit{Upmanyu et al} 
use a so-called shatter function (a function described \cite{Upmanyu:2010:EPP:2163922.2163944} to divide a value into many secret shares keeping the privacy of the data intact) and the Chinese Remainder theorem \cite{goldreich1999chinese,mignotte1982share} to encrypt and reconstruct respectively. They propose a ‘cloud computing’ based solution that utilizes the services of  non-colluding servers. Each of the  users, is required to compute the  secret shares of its private data using a shatter function. Each share is then sent over to a specific server for processing. The cloud of employed servers, now runs the $k$-means algorithm using just the secret shares. The protocol ensures that none of the users/servers have sufficient information to reconstruct the original data, thus ensuring privacy.

\subsection{Our Contribution}
\label{S:1:1}
We use the concept of outsourcing~\cite{Liu:2014:POK:2590296.2590332} the data to third parties who will do the computation for the data provider. These third parties are considered as adversarial, hence the data needs to be protected from them. Though \cite{Upmanyu:2010:EPP:2163922.2163944} is fairly efficient, our protocol is better because we avoid any cryptographic overheads and use multiplicative data perturbation. Since our protocol divides the data into parts and every server works in parallel, it boosts the performance in comparison to a single server performing the whole algorithm  \cite{Yu2010MultipartyKC}. We argue that our protocol is secure against attacks on data perturbations  because of the introduction of a noise term. Keeping the noise under a certain limit, we have been able to provide a clustering algorithm that achieves the same accuracy as the iterative $k$-means over 
non-randomized data.

\section{Proposed Solution}
\label{S:2}
\subsection{Problem Setup}
\label{S:2:1}
In our setting, there is one data owner who holds a large dataset $\mathcal{D}$ containing $n$ data points each having $d$ attributes. All the attributes are considered to be floating point parameters. Hence $\mathcal{D}$ can be thought as containing
$n$ points in $\mathcal{R}^d$. Let these points be labeled as $X_{1}, X_{2},...,X_{n}.$ 

The data owner wishes to use $t$ servers to compute the $k$-means. In this work we consider horizontal partitioning of the data,
which means dividing the entire dataset into subsets based on tuples. Each tuple contains all the attributes involved. 
Let $m$ be the number of iterations needed for the $k$-means to converge where $k$ represents the number of clusters we want to 
form. The problem is computing the $k$-means on the entire dataset securely, efficiently and accurately by dividing the dataset horizontally among the servers without revealing any information about the original data points and any of the attributes to the servers.

\subsection{Our Protocol}
\label{S:2:2}
\begin{itemize}
\item The data provider generates $2d$ random numbers $r_{i}, i\in{1,2,...,2d}$ from a large set $\mathbb{R}$.
A lower bound for the value of $r_i$ will be discussed in a later section.
\item The data provider selects a small enough $\epsilon>0$ and then chooses $n$ many $\epsilon_{i}$, $i\in{1,2,...,n}$ uniformly from $(0,\epsilon)$. They will behave as noise added to the data to improve the security. A detailed analysis of the upper bound of 
$\epsilon$ has been provided later.
\item Let us denote $X_{i}= (x_{i1},x_{i2},...,x_{id})$. \\
Randomize the data by the following computation:\\ 
\begin{equation}
\label{e1}
X_i'=((r_1+\epsilon_i)*x_{i1}+r_2,(r_3+\epsilon_i)*x_{i2}+r_4,...,(r_{2d-1}+\epsilon_i)*x_{id}+r_{2d}).
\end{equation}
Hence the $j^{th}$ attribute of $X_i$ is transformed to: 
\begin{equation}
\label{e2}
(r_{2j-1}+\epsilon_i)*x_{ij}+r_{2j}.
\end{equation}
\item The data owner then locally partitions the transformed data horizontally into $t-1$  parts and sends it to $t-1$ servers, which means the $t^{th}$ server does not receive any data. The work of the $t^{th}$ server shall be to perform certain calculations using the data provided to it by the remaining servers. The details of which shall be discussed below.
\end{itemize}

\newpage
\subsubsection{The $k$-mean computation}
\label{S:2:2:1}

\paragraph{Initialization Step}
The data provider picks $k$ many transformed data points at random.\footnote{We are aware of several other methods to select the initial centers which may make the $k$-means work more efficiently. But in this work we do not concentrate on assignment of initial clusters too much.} These points will act as the initial cluster centers. These points say, $c_1,c_2,...,c_k$ shall be sent to all the $t-1$ servers who have some part of the transformed data points.

\paragraph{Lloyd's Step}
\begin{enumerate}
\item Each server computes the Euclidean distance of its share of data from the initial centers and assigns cluster labels to the points locally.
\item Every server finds the number of points allotted to each center among their share of the data. Suppose for server $s_i$,
$m_{ij}$ denotes the number of data points belonging to cluster $c_j$. Here $i \in \{1,2,...,(t-1)\}, j \in \{1,2,...,k\}.$
\item Each server computes the sum of the points belonging to each center. Let us denote it by $d_{ij}$, which denotes the sum of the points belonging to cluster $j$ for server $i$.
\item Next step involves the generation and sharing of two secret keys $x$ and $y$ among the $t-1$ servers. For this purpose, the data owner may generate two random numbers and transfer them to the $t-1$ servers alongside the transformed data set that is being transferred. The key sharing will be performed only for the first iteration. From the next iteration onwards we can use a
cryptographically secure hash function to get modified values of $x$ and $y$ for every step. The hash function used will be a common function known to each of the first $t-1$ servers. The key generation procedure and hash function will be discussed in details
in \S\ref{S:2:3}.
\item Each server computes 
\begin{equation}
\label{e3}
x*d_{ij}
\end{equation}
and 
\begin{equation}
\label{e4}
y*m_{ij}
\end{equation}
for each center $j$, $1 \leq j \leq k$ and sends it to the $t^{th}$ server. 
\item Server $t$ calculates:
\begin{equation}
\label{e5}
\frac{\sum_{i=1}^{t-1}(x*d_{ij})}{\sum_{i=1}^{t-i}(y*m_{ij})}
\end{equation}
for $1 \le j \le k$ centers and returns this result to the other servers. This value shall work as the centroid i.e. the new centers for the subsequent iteration of the Lloyd's step. \\
Let the centroid be denoted by $\nu_1,\nu_2,...,\nu_k$.
\end{enumerate}

\paragraph{Re-initialization}
Repeat Lloyd's step till convergence.
If the new centroids computed are not equal to the centroids computed in the previous iteration, i.e.
$\{c_1,c_2,...,c_k\} \neq \{\nu_1,\nu_2,...,\nu_k\}$ then reassign $c_1,c_2,...,c_k=\nu_1,\nu_2,...,\nu_k$\\. 
These shall be the updated centroid values.

\paragraph{Output}
After the iterations are complete, the $t-1$ servers send the cluster centers and cluster assignments of their share of data to the data provider. The data owner now possesses the cluster labelling of all the data and the final cluster centers. Hence, the algorithm terminates at this step.

\subsection{Group Key Sharing and Hash Function}
\label{S:2:3}
As previously mentioned, the initial set of $x$ and $y$ will be provided to the $t-1$ servers by the data provider. Alternatively the $t-1$ servers may indulge in a group key sharing algorithm~\cite{cryptoeprint:2018:114} to generate the first pair of random numbers. But this would lead to additional computational costs which we are compensating for, by a little bit of additional communication   that is involved in transferring two random numbers from the data owner to servers. For the subsequent iterations, we use a publicly
available hash function. This function will take as input the output of the previous iteration and the round number. This will allow only the parties that have access to the group key to generate random numbers using the hash iteratively. The hash function used
should be a one way function, i.e. computation of the inverse of the hash function should be computationally hard.

\subsection{Dynamic Setting}
\label{S:2:4}
In our protocol, we have only talked about static data. In case, the data provider gets access to more data that it wishes to include in the $k$-means calculation, the data provider does the randomization as equation (1) over the new data points. The new points will then be partitioned and sent to the servers. These servers will just include these new points during the assignment of clusters and finding of centroids from the subsequent iteration and proceed as before till convergence.

\section{Analysis}
\label{S:3}
In this section, we will be performing a detailed analysis of the correctness of our protocol. We will inspect the accuracy and show how we handle error in our protocol by providing an upper bound for the noise element. Further we will scrutinize our protocol from the security point of view, where we will talk about leakage of information and conclude how such a leakage does not compromise the privacy. Lastly we will provide a brief account of the efficiency of our algorithm.

\subsection{Correctness}
\label{S:3:1}
Since iterative $k$-means guarantees convergence, our protocol will be deemed correct if we can prove convergence of our algorithm over the transformed data and if we can show that the error involved in clustering the transformed data is acceptable when compared to the clustering of the original data.

Without loss of generality, we have made the assumption that all data points have non negative attributes. This assumption can be made because all points can easily be translated such that all their coordinates become positive. This is done without distorting the geometry at all, hence it does not affect the clustering algorithm.

Necessarily the main operations in the $k$-means computation are the following steps.
\begin{enumerate}
\item Find Distance: Computing distance between points and the centroids.
\item Compare Distance: Find which centroid is nearest to a point.
\item Find new centroid: Re-initialize the centers.
\end{enumerate}
For two data points $X_1, X_2$ and the corresponding transformed data points $X_1', X_2'$ we can express the respective distances as follows.
\begin{equation}
\label{e6}
\mbox{Distance}(X_1,X_2) := \sqrt[2]{\sum_{i=1}^d{(x_{1i}-x_{2i})^2}}
\end{equation}

\begin{eqnarray}
\label{e7}
& &\mbox{Distance}(X_1',X_2') := \sqrt[2]{\sum_{i=1}^d{(r_{2i-1}(x_{1i}-x_{2i})+\epsilon_1x_{1i}-\epsilon_2x_{2i})^2}}\\
&=& \sqrt[2]{\sum_{i=1}^d({r_{2i-1}^2(x_{1i}-x_{2i})^2+(\epsilon_1x_{1i}-\epsilon_2x_{2i})^2+2r_{2i-1}(x_{1i}-x_{2i})(\epsilon_1x_{1i}-\epsilon_2x_{2i}))}}.\nonumber
\end{eqnarray}
In order to make our calculations and analysis simpler, the above expression under the root maybe looked upon as a quadratic polynomial in $\epsilon$. Assuming $\epsilon$ to be sufficiently small (say $\leq 0.1$), the polynomial will be dominated by the lower order and the constant terms, hence the quadratic term of $\epsilon$ can be neglected. Thus, the final expression takes the form:
\begin{equation}
\label{e8}
\sqrt[2]{\sum_{i=1}^d({r_{2i-1}^2(x_{1i}-x_{2i})^2+2r_{2i-1}(x_{1i}-x_{2i})(\epsilon_1x_{1i}-\epsilon_2x_{2i}})}.
\end{equation}
In some places we will use expression (\ref{e8}) instead of (\ref{e7}) and will provide proper justification for its usage.
We introduce an error term $\lambda$ where $\lambda_1=(\epsilon_1x_{1i}-\epsilon_2x_{2i})^2+2r_{2i-1}(x_{1i}-x_{2i})(\epsilon_1x_{1i}-\epsilon_2x_{2i})$ and $\lambda_2=2r_{2i-1}(x_{1i}-x_{2i})(\epsilon_1x_{1i}-\epsilon_2x_{2i})$ shall be considered to be the error terms  for (\ref{e7}) and (\ref{e8}) respectively added to the distance due to inclusion of noise. Note that the error terms
$\lambda_1, \lambda_2$ are directly proportional to $\epsilon$ and the difference in co-ordinates of the two points ($X_1,X_2$). Given a sufficiently small $\epsilon$, the error can be easily bounded by an acceptable threshold.
This will be made clearer shortly when we discuss the bound of $\epsilon$ and $r_i$s. We emphasise that
the distance between the transformed points $(X_1',X_2')$ is nothing but the scaled distance between the original points
$(X_1,X_2)$ with some error term added to it. The scaling is done uniformly for all data points $X_i$, $1 \le i \le n$.

It is evident that $k$-means converges when the distance between the new and original centroid becomes $0$. 
Expression (\ref{e6}) becomes $0$ when $x_{1i}-x_{2i}=0, \forall i\in 1,...,d$. Given that each $r_i>0$, it will be evident that 
(\ref{e8}) will become $0$ if and only if the above condition holds. Since our exact distance form is described by (\ref{e7}), (\ref{e8}) becoming $0$ implies an error element may prevail in (\ref{e7}) that may not become $0$, but we can neglect that error under the above assumption of sufficiently small $\epsilon$. 

Hence, we can claim that the $k$-means on the transformed data shall converge at the same time when the $k$-means on the plain text converges. So, the number of iterations required for convergence is exactly the same.

\subsubsection{Lower bound of $r_i$}
\label{S:3:1:1}
We need to specify a range of $r_is$ for which the error term involved in the above expression can be acceptable.
We can say that the error term will not influence our clustering if it does not alter our Compare Distance method. 
Hence, if 
\begin{equation}
\label{e9}
\sqrt[2]{\sum_{i=1}^d{(x_{1i}-x_{2i})^2}}<\sqrt[2]{\sum_{i=1}^d{(x_{1i}-x_{3i})^2}}
\end{equation}
then,
$$\sqrt[2]{\sum_{i=1}^d({r_{2i-1}^2(x_{1i}-x_{2i})^2+(\epsilon_1x_{1i}-\epsilon_2x_{2i})^2+2r_{2i-1}(x_{1i}-x_{2i})(\epsilon_1x_{1i}-\epsilon_2x_{2i}))}}$$
\begin{equation}
\label{e10}
<\sqrt[2]{\sum_{i=1}^d({r_{2i-1}^2(x_{1i}-x_{3i})^2+(\epsilon_1x_{1i}-\epsilon_3x_{3i})^2+2r_{2i-1}(x_{1i}-x_{3i})(\epsilon_1x_{1i}-\epsilon_3x_{3i}))}}
\end{equation}
for all possible values of $i$, which means,
$$\mbox{max}\sqrt[2]{\sum_{i=1}^d({r_{2i-1}^2(x_{1i}-x_{2i})^2+(\epsilon_1x_{1i}-\epsilon_2x_{2i})^2+2r_{2i-1}(x_{1i}-x_{2i})(\epsilon_1x_{1i}-\epsilon_2x_{2i}))}}$$
$$<\mbox{min}\sqrt[2]{\sum_{i=1}^d({r_{2i-1}^2(x_{1i}-x_{3i})^2+(\epsilon_1x_{1i}-\epsilon_3x_{3i})^2+2r_{2i-1}(x_{1i}-x_{3i})(\epsilon_1x_{1i}-\epsilon_3x_{3i}))}}$$
Solving the above equation with proper bounds and using the expression (\ref{e10}) we get a lower bound for $r_is$.
\begin{equation}
\label{e11}
r> max\frac{\sum_{l=1}^d{(x_{il}^2-(x_{il}-x_{kl})^2)}}{2 \sum_{l=1}^d{(x_{il}-x_{kl})^2-x_{il}(x_{il}-x_{jl})}}, \forall i, j, k
\end{equation}
where r=$\mbox{min}(r_i), \forall i$. Refer to Appendix \ref{A:1} for detailed calculation.

\subsubsection{Upper Bound on $\epsilon$}
\label{S:3:1:2}
The requirement that the term inside the root in expression (\ref{e8}) must be non-negative, gives us an upper bound for 
$\epsilon$. A sufficient condition to achieve it is,
$r_{2i-1}^2(x_{1i}-x_{2i})^2>\lambda_2$. Substituting for the value of $\lambda_2$
we get, 
$r_{2i-1}(x_{1i}-x_{2i})>2(\epsilon_1x_{1i}-\epsilon_2x_{2i})$ over all possible values of $i$.
Hence, $r_{2i-1}(x_{1i}-x_{2i})>\mbox{max}(2(\epsilon_1x_{1i}-\epsilon_2x_{2i}))$.
Upon simplification, we get
\begin{equation}
\label{e12}
\epsilon< \mbox{min} \frac{r_{2k-1}(x_{ik}-x_{jk})}{2x_{ik}}, \forall i,j,k.
\end{equation}
If we use (\ref{e7}) as our parent equation then (\ref{e9}) remains unchanged. This is so because $\lambda_1$ is greater than 
$\lambda_2$. So using the latter provides us a stronger upper bound for $\epsilon$.

Equations (\ref{e11}) and (\ref{e12}) show the bound of $r$ and $\epsilon$. 
Combining the two relations, we will get a common expression for the relation between $\epsilon$ and $r$ that shall ensure correctness. Thus if $\epsilon$ and $r_i$s lie in this range then the output of Compare Distance function will not be altered 
for the transformed data.

It is guaranteed that assignment of intermediate clusters for the points remain consistent with the assignment without the transformation. That is so because the centroids are found by taking the average over the points in a particular cluster, hence the distance between a centroid and a data point will be less than the global maximum and more than the global minimum as described in the derivation of (\ref{e11}) and (\ref{e12}) respectively. These conditions will ensure that the clustering over the randomized 
data points is same as the clustering over the original data. This way we ensure the accuracy of our clustering technique.

\subsection{Security}
\label{S:3:2}
We consider the adversarial servers to be honest-but-curious.
\paragraph{Adversarial Power:}
\begin{enumerate}
\item Every server tries to obtain maximum information about the original data without deviating from the protocol.
\item Every server would like to gain knowledge about the data possessed by the other servers.
\item Servers record and store all intermediate information made available to them and use it to find out more information about the data.
\item Collusion among servers is not allowed.
\end{enumerate}
\paragraph{Information available to servers $1$ to $t-1$:}
\begin{enumerate}
\item Randomized data points.
\item Intermediate cluster assignments of their own data only.
\item Intermediate cluster centers
\item Number of iterations needed to converge.
\end{enumerate}
\paragraph{Information available to $t^{\mbox{th}}$ server:}
\begin{enumerate}
\item A scaled version of the intermediate centers.
\item Randomized sum of coordinates of the data points that belong to a particular cluster at each iteration.
\item Randomized value for the number of data points belonging to every cluster for each server at every iteration.
\item Randomized value of intermediate centers.
\item Number of iterations needed to converge.
\end{enumerate}

\subsubsection{Security Against Existing Attack Scenarios}
\label{S:3:2:1}
Recall that the information initially available to the first $t-1$ servers is of the form of Equation (1). 
Various algebraic methods have been discussed in \cite{Kargupta2005,10.1007/11871637_30,Liu2006RandomPM} 
to design attacks on data perturbation techniques. But most of the attacks described so far in the above works
are applicable for additive noise. In \cite{10.1007/11871637_30}, Liu et al have discussed in detail the
security in random perturbation from the attackers point of view. Their model deals with known sample or known 
input-output models in case of Distance Preserving Transformations. Liu et al \cite{10026864614} talked about attacks on multiplicative data perturbation. Their approach uses Independent Component Analysis to remove the randomization and gain information about the data. Given that Principle Component Analysis (PCA) works successfully only when the perturbation matrix is 
orthogonal, so if the transformation is not distance preserving as in our case, PCA is unsuccessful to gain any significant information about the original data.

The crux of all these attack approaches is the fact that that distance preserving transformation in a vector space over a real field is an orthogonal transformation. The advantage of our technique is that the data transformation does not preserve the distance hence making the transformation a non-orthogonal one. This fact appears to make it more secure than the Distance Preserving transformations. 

A recent work \cite{kaplan2017known} has devised an attack on Relation Preserving Transformation (RPT). 
Since RPT is the basis of our transformation, it may be open to breach by \cite{kaplan2017known}. Assuming that such an attack is implemented, we analyze the feasibility of it in detail.

We first recall the salient features of the attack proposed in \cite{kaplan2017known}.
It is assumed that there exists a third party malicious adversary and the attacker has knowledge about some original data points.
The attack reveals which side of the hyper-plane does the point lie. No information is found about the exact location of the point. A major assumption is that the search space is discrete. It has been stated that the algorithm is useful for data set that is usually low dimensional. The main basis of a successful attack is that probability of choosing a point inside a bounded area is non negligible which again goes back to the assumption of a discrete search space.

As per \cite{kaplan2017known}, the complexity of their algorithm is $O({|K|\choose2} {(\frac{R}{c})}^d \mathcal{I})$. 
In the above expression, $|K|$ is the size of known sample, $R$ is the range of data points,
$c$ the length of a cell into which the entire space is divided and $\mathcal{I}$ is the complexity of finding Intersection.

We argue why such an attack is not practically applicable in our case.
The attacker has no knowledge about any of the original data and
all communication channels are assumed to be secure and the servers have no information about
any data point. We are working with high dimensional data sets where, usually $d \ge 8$. 
We consider the best case scenario for the attacker and take $|K|=2$. In the following table, we tabulate complexity of the attack for different choices of $R, c$ and $d$.

\begin{table}
\label{Tab:1}
\begin{center}
\begin{tabular}{|l|l|l|l|}
\hline
$R$ & $c$ & $d$ & ${x}$ in $O({x}\mathcal{I})$\\ 
\hline
1000 & 0.01&2&$2^{33}$\\
1000	& 0.01 & 3 & $2^{49}$\\
1000&0.01&4&$2^{66}$\\
10&0.001&5&$2^{66}$\\
10&0.001&6&$2^{79}$\\
10&0.001&8&$2^{105}$\\
100&0.01&9&$2^{118}$\\
10&0.001&10&$2^{132}$\\
10&0.001&12&$2^{158}$\\
1000&0.01&11&$2^{181}$\\
\hline
\end{tabular}
\caption{Complexity of the attack ($O({x}\mathcal{I})$), where $x= {|K|\choose2} {(\frac{R}{c})}^d$ from \cite{kaplan2017known} in our setting.}
\end{center}

\end{table}

In the current computational power, around $2^{64}$ steps is considered barely feasible. To compute $k$-means, let's assume
we require precision of at least 3 digits for accuracy. Since we are dealing with large values of $d$, even in the best case for the attacker where $R=10$, the complexity can be seen to be much larger than $2^{64}$. While dealing with large datasets, it is not a practical assumption that the data points are dispersed over a range of just 10 units. It will be much more than this in most cases and $d \geq 8$ in most cases where cloud computing is used. We can thus conclude from the above table that this attack cannot be practically implemented whenever the dimension is more than $5$ because of the extremely high complexity. Since the attack is exponential in $d$, the attack becomes extremely inefficient for large and high dimensional data sets making it infeasible to implement in real life.

\subsubsection{Security Against Data Leakage}
\label{S:3:2:2}
Our technique allows certain information leakage to the servers. After convergence, servers $1$ to $(t-1)$ will get to know
information about the final cluster allotments for every data-point that they have access to. They also learn
which points belong to the same cluster. The $t^{\mbox{th}}$ servers knows the intermediate as well as the final cluster centres.
This leakage allows the servers to gain information about the transformed points only. Knowing about the cluster assignments of the randomized points does not help the adversary gain any significant information about the original points or their cluster assignments. We shall now justify this statement with concrete analysis.

The adversarial servers may try to remove the randomness from the data they have and retrieve maximum information about the original data. If they take the attribute wise quotient of their data then they get the following:
\begin{equation}
\label{e13}
\frac{r_1(x_{11}-x_{21})+\epsilon_1x_{11}-\epsilon_2x_{21}}{r_1(x_{41}-x_{31})+\epsilon_4x_{41}-\epsilon_3x_{31}}.
\end{equation}
If the servers wish to use the entire data points instead of the attributes, then the one possible method to proceed will be to compute the generalized inverse\cite{mitra1968generalised} by treating the vectors as column matrix. Finding the g-inverse of a point and multiplying it with another data point can be interpreted as a quotient between two vectors. This calculation leads us back to a form of the above expression (\ref{e13}). 

We next analyse the effectiveness of a probabilistic approach to see if there is some significant leakage of data. 
We want to ensure that the above expression (\ref{e13}) reveals no significant information about
\begin{equation}
\label{e14}
\frac{x_{11}-x_{21}}{x_{41}-x_{31}}.
\end{equation}
We assume probability distributions over expression (\ref{e13}) and (\ref{e14}) and proceed to check how similar are these two distributions. If the distributions are not similar then we can claim the expression (\ref{e13}) does not reveal anything 
significant about expression (\ref{e14}).

We use the Kullback-Leibler Divergence function \cite{hershey2007approximating} as a metric to compare the two distributions. Kullback-Leibler divergence is a bounded function between $0$ and $1$. The further the value is from $0$, the less similar 
are the two distributions.
With the help of proper upper and lower bounds, simplification of the divergence functions gives us a lower bound on the metric. Let us denote KD as the output of the divergence function. Then,
\begin{equation}
\label{e15}
KD \geq  d\frac{x_{11}-x_{21}}{x_{41}-x_{31}} \log{ \frac{r_1+\frac{\epsilon x_{41}}{x_{41}-x_{31}}}{r_1-\frac{\epsilon x_{11}}{x_{11}-x_{21}}}}.
\end{equation}
Refer to Appendix \ref{A:2} for details.

The definition of Kullback-Leibler guarantees the value of (\ref{e15}) to be non negative.  Since (\ref{e15}) is an increasing function of $\epsilon$, the greater the value of $\epsilon$, more is the deviation of the function from $0$. We can increase 
$\epsilon$ till the upper bounds to ensure that the Kullback-Leibler distance moves away from $0$.  Hence, by regulating $\epsilon$, the probability distributions can be made dissimilar.

Finally, we discuss the leakage of information to server $t$. Server $t$ receives information in the form of equations 
(\ref{e3}) and (\ref{e4}). Its aim again will be to remove the randomization and get information about the original values. 
It can try the following two divisions to extract out the randomness.
Compute
\begin{equation}
\label{e16}
\frac{(x*d_{ij})}{(y*m_{ij})}
\end{equation}
or compute using only (\ref{e3})
\begin{equation}
\label{e17}
\frac{(x_1*d_{ij})}{(x_2*d_{ij})}
\end{equation}
and do similar with the use of (\ref{e4}) alone.
Again, using the same techniques as before of assuming probability distributions and finding the Kullback-Leibler divergence 
function between the randomized and the non-randomized values, it can be shown that KD for (\ref{e16}) is:
\begin{equation}
\label{e18}
-\log{\frac{x}{y}}\sum{\frac{d_{ij}}{m_{ij}}}
\end{equation}
while KD for (\ref{e17}) is:
\begin{equation}
\label{e19}
-\log{\frac{x_1}{x_2}}z
\end{equation}
where $z$ denotes the number of points taken into consideration while computing KD.

Thus we see that as long as $x$ and $y$ are not same, the occurrence of which has negligible probability as the numbers are 
generated randomly, the Kullback-Leibler divergence function will give an output that will be away from $0$.

There is no interaction between servers $1,...,(t-1)$ other than the key exchange, so a server cannot gain any information about the data of the other servers when collusion is disallowed. The other leakage of information that we compromise with is the number of iterations needed to converge, but we can accommodate this because it does not give up on the privacy of the data which is our primary goal.

\subsection{Efficiency}
\label{S:3:3}

\paragraph{}
In analyzing the performance of our algorithm on the basis of the total communication and computational cost, we discuss the complexity of the entire process by dividing it into three different stages: the data provider, the first $t-1$ servers and 
the $t^{th}$ server.

\paragraph{Data Provider}
Computation: The only computation done here is the randomization of the data where the computation cost is dominated by the number of multiplications.

Communication: There shall be a one time communication necessary to send the randomized data to the respective servers. Without loss of generality, we may assume that the data provider divides the data set into $t-1$ parts each of size 
$n_1, n_2,...,n_{t-1}$. The communication cost will depend on the size of the data transferred. In this and all further cases that we discuss, we will deal with the worst case, i.e. we assume that the size of the data is the upper bound for all the possible values. Say this upper bound is $U$.

\paragraph{Servers $1$ to $(t-1)$}
Computation: The main computations being done here are finding distance and comparing distance before assigning the necessary clusters. Here the operations that dominate the performance are performing squares and doing comparisons to find which cluster a point should belong to.

Communication: Sending $x*d_{i,j}$ and $y*m_{i,j}$ (see expressions (\ref{e3}) and (\ref{e4}) respectively) to server $t$ uses up some bandwidth. Here consider that all values sent by a server $i$ in the form of (\ref{e3}) and (\ref{e4}) shall have an upper bound $N_i$ and $M_i$. This communication cost will be accounted for $m$ number of times where $m$ is the number of iterations needed for the algorithm to converge.

\paragraph{Server $t$}
Computation: Computing the intermediate cluster centres (see (\ref{e5})) requires division operation that will be the main
computational cost in this case.

Communication: Returns $k$ many values of intermediate cluster centres to each server $m$ number of times. 
Assumption is that the value of those centres will always be less than $C$, where $C$ is the upper bound of 
all values to be returned by server $t$.\\
 
\paragraph{Comparison with \cite{Upmanyu:2010:EPP:2163922.2163944}}
Since our model is closest to the one proposed by {Upmanyu et al}, it is fair to compare the efficiency of both the approaches. Instead of using three layers of interaction like us, they use only two levels of interaction. Their communication cost for sending data from data owner to servers is same as ours because they have to send secret shares of each data point to the servers similar to our sharing of data points to servers. Computationally, our algorithm beats theirs because of the following. (i) In \cite{Upmanyu:2010:EPP:2163922.2163944}, the data owner needs to shatter the data points leading to performing $t$ modulo operations for each data point. Hence, $nt$ modulo operations are to be performed, whose computation cost is similar to inversion that is heavier than multiplication. (ii) At the server level, in order to assign clusters, the servers need to merge their share of secrets together and then proceed with distance computation and comparison. This process in whole involves two main operations, sharing common secret keys using group key sharing and merging the shared secrets. The merging operation uses Chinese Remainder Theorem (CRT), which has a complexity of O($N^2$), where $N$ is the modulus in CRT. Repeating this for all $n$ data points and for $m$ many rounds makes the complexity $O(nN^2m)$. In addition to this, the usage of a common group key sharing algorithm further increases the computation cost significantly. Thus the amount of computations to be done by \cite {Upmanyu:2010:EPP:2163922.2163944} is much heavier than in our case. We summarize in the tables below, the comparison between computation cost of the two algorithms.

\begin{table}
\label{T:2}
\begin{center}
\begin{tabular}{|c|c|c|}
\hline
 & Computational & Communication\\ 
 \hline
 Data Provider & $O(nd)$(Multiplication) & $O(nU)$  \\  
 Servers $1$ to $(t-1)$ & $O(kn_idm)$(Multiplication) & $O(m(N_i+M_i))$ \\
 Server $t$ & $O(mk)$(Inversion) & $O(mkC)$\\
\hline
\end{tabular}
\caption{Our Algorithm}
\end{center}
\end{table}

\begin{table}
\label{T:3}
\begin{center}
\begin{tabular}{|c|c|c|}
\hline
 & Computational & Communication\\ 
 \hline
 Data Provider & $O(nt)$ (Inversion) & $O(nU)$  \\  
 Servers & $O(nN^2m)$ (Chinese Remainder Theorem) & $O(n_iN_i)$\\
\hline
\end{tabular}
\caption{Algorithm in  \cite{Upmanyu:2010:EPP:2163922.2163944}}
\end{center}
\end{table}

\subsubsection{Performance Comparison when data owner locally compute $k$-means}
\label{S:3:3:1}
If the data provider did not outsource the computation of $k$-means and instead did the entire process on his/her own, 
the complexity would be $O(nkdm)$. The performance would be dominated by multiplications and inversions. Whenever the number of clusters to be formed becomes large, the efficiency would be affected. By outsourcing, one will also be relieved of performing numerous inversions and comparisons that will be taken care of by the servers. Moreover, along with time complexity, another constraint might be space complexity as well. If the entire algorithm is performed locally, then the data owner needs storage space in order to keep all the intermediate information recorded at every round of iteration. In this case, all that
the data provider needs is storage for the data set for only the first round. 

\section{Choice of Parameters for Practical Implementation}
\label{S:4}
\paragraph{}
It may seem that pre-processing the data for randomization will require a significant amount of computation.
Following the naive approach, the first step would be selecting values of $r_i$ and $\epsilon$. Following (\ref{e11}),
to find the strict bound of $r$ shall take O($n^3$) many inversions. This would go against our claim of having a very efficient 
algorithm. However, the problem can be easily solved by using a weaker bound instead of using the strict bound that we have derived in (\ref{e11}) and (\ref{e12}). From (\ref{e11}) we have,
$$r> \mbox{max}\frac{\sum_{l=1}^d{(x_{il}^2-(x_{il}-x_{kl})^2)}}{2 \sum_{l=1}^d{(x_{il}-x_{kl})^2-x_{il}(x_{il}-x_{jl})}}, \forall i, j, k.$$
Note that,
$$\mbox{max}\frac{\sum_{l=1}^d{(x_{il}^2-(x_{il}-x_{kl})^2)}}{2 \sum_{l=1}^d0{(x_{il}-x_{kl})^2-x_{il}(x_{il}-x_{jl})}}
\leq \mbox{max}\frac{\sum_{l=1}^d{(x_{il}^2-(x_{il}-x_{kl})^2)}}{2 \sum_{l=1}^d{(x_{il}-x_{kl})^2-x_{il}^2)}}
\leq -\frac{1}{2}.$$
Since (\ref{e11}) gives the range for correctness, $r>-\frac{1}{2}$ will retain the correctness.
From (\ref{e12}), we have,
$$\mbox{min}\frac{r_{2k-1}(x_{ik}-x_{jk})}{2x_{ik}} \geq \mbox{min}{\frac{r_{2k-1}}{2}}.$$
Since we are dealing with only positive values of $\epsilon$, one can choose any non-negative real number $w$. Then choosing 
$r>w$ and $\epsilon < \frac{w}{2}$ will ensure correctness.

This process helps us get the value of the parameters in constant time. The next step would be performing multiplications to randomize the data. Our aim is to optimize security and efficiency. We use the bit length of $r_i$ and $\epsilon_i$ to analyze the efficiency and the security. The efficiency is dominated by the multiplications to be performed. Multiplying two numbers of 
$\ell$-bits has a complexity of $O(\ell^2)$. Total $nd$ many multiplications are needed to be performed that will be a complexity 
of $O(nd\ell^2)$. Let the bit length of the maximum value of $r_i$ be $\ell_1$ and that of the maximum value of $\epsilon_i$ be 
$\ell_2$.
We assume that our algorithm is deemed secure if the adversary cannot guess the random numbers with probability more that 
$2^{-80}$. We analyze the security of two expressions. In the first, the adversary needs to guess two values of 
$r_i$s and one value of $\epsilon_i$ to get to know about one of the coordinates of a data point from expression~\ref{e1}:
$$X_i'=((r_1+\epsilon_i)*x_{i1}+r_2,(r_3+\epsilon_i)*x_{i2}+r_4,...,(r_{2d-1}+\epsilon_i)*x_{id}+r_{2d}).$$
For the second case, the adversary has to guess one value of $r_i$ and three values of $\epsilon_i$s with non negligible probability from equation \ref{e13} as reproduced below:
$$\frac{r_1(x_{11}-x_{21})+\epsilon_1x_{11}-\epsilon_2x_{21}}{r_1(x_{41}-x_{31})+\epsilon_4x_{41}-\epsilon_3x_{31}}.$$
In the following table we demonstrate some plausible values of $\ell_1$, $\ell_2$ that will optimize security along with correctness. The way of choosing $\ell_1$ and $\ell_2$ has been talked about in details in Appendix \ref{B}.\\
We consider $n= 2^{16}$ and $d=2^4$. One assumption is that $\ell_1 > \ell_2$ as we do not want the noise to surpass the scaling factor.

\begin{table}
\label{T:4}
\begin{center}
\begin{tabular}{|c|c|c|c|c|}
\hline
 $\ell_1$& $\ell_2$ & Probability of guessing (1) &  Probability of guessing (13) & $O(nd\ell_1^2)$\\ 
 \hline
34 & 32 & $2^{-77}$ & $2^{-80}$ & $2^{30}$\\
40 & 32 & $2^{-89}$ & $2^{-82}$ & $2^{31}$\\
64 & 32 & $2^{-137}$ & $2^{-110}$ & $2^{32}$\\
128 & 8 & $2^{-241}$ & $2^{-102}$ & $2^{34}$\\
\hline
\end{tabular}
\caption{Parameters for Practical Implementation}
\end{center}
\end{table}

\section{Conclusion}
\label{S:5}
In this work, we have proposed a solution to perform cloud-based $k$-means clustering in the multi-server setting. 
The main aim was to perform clustering as efficiently as possible without compromising with the privacy of the data to the
extent possible. We have provided a technique that is easy to understand and implement along with being robust. In our work, 
we have analyzed the correctness and security of the algorithm in details. Our analysis shows that the
proposed technique is secure against a passive adversary. Our method is very efficient as it does not include any heavy
cryptographic computation. The $k$-means process we have described is similar to the iterative $k$-means used over original 
data set. Hence the efficiency of both the algorithms is comparable. We also discussed practical parameter choices for our 
algorithm. One interesting future work would be to extend the perturbation based approach to allow partial collusion
among the servers.

\bibliographystyle{splncs04}

\appendix
\section{Detailed Computations}
\label{A}
\subsection{Lower Bound for $r_i$, $1 \le i \le 2d$:}
\label{A:1}
\begin{eqnarray*}
& & \mbox{max}\sqrt[2]{\sum_{i=1}^d({r_{2i-1}^2(x_{1i}-x_{2i})^2+(\epsilon_1x_{1i}-\epsilon_2x_{2i})^2+2r_{2i-1}(x_{1i}-x_{2i})(\epsilon_1x_{1i}-\epsilon_2x_{2i}))}} \\
&<& \mbox{min} \sqrt[2]{\sum_{i=1}^d({r_{2i-1}^2(x_{1i}-x_{3i})^2+(\epsilon_1x_{1i}-\epsilon_3x_{3i})^2+2r_{2i-1}(x_{1i}-x_{3i})(\epsilon_1x_{1i}-\epsilon_3x_{3i}))}}
\end{eqnarray*}
To maximize LHS and minimize RHS, we take $\epsilon_1$=$\epsilon$, $\epsilon_2$=0, $\epsilon_3=\epsilon$ and thus,
\begin{eqnarray*}
& & \sum_{i=1}^d({r_{2i-1}^2(x_{1i}-x_{2i})^2+(\epsilon^2x_{1i}^2+2r_{2i-1}(x_{1i}-x_{2i})(\epsilon x_{1i}))}\\
&<& \sum_{i=1}^d({r_{2i-1}^2(x_{1i}-x_{3i})^2+(\epsilon^2(x_{1i}-x_{3i})^2+2r_{2i-1} \epsilon (x_{1i}-x_{3i})^2)}
\end{eqnarray*}
Using (10), we further get,
\begin{eqnarray}
& & \epsilon \sum_{i=1}^d{[x_{1i}^2-(x_{1i}-x_{3i})^2]}+2 \sum_{i=1}^d{r_{2i-1}[x_{1i}(x_{1i}-x_{2i})-(x_{1i}-x_{3i})^2]}<0\nonumber\\
&\Rightarrow& 2 \sum_{i=1}^d{r_{2i-1}[(x_{1i}-x_{3i})^2-x_{1i}(x_{1i}-x_{2i})]}>\epsilon \sum_{i=1}^d{[x_{1i}^2-(x_{1i}-x_{3i})^2]}\nonumber \\
\label{e20}
&\Rightarrow& r> \mbox{max} (\epsilon \frac{\sum_{l=1}^d{(x_{il}^2-(x_{il}-x_{kl})^2)}}{2 \sum_{l=1}^d{(x_{il}-x_{kl})^2-x_{il}(x_{ik}-x_{jl})}}), \forall i, j, k.
\end{eqnarray}
Given that $\epsilon$ is sufficiently small, it can be safely assumed to be less than 1. Hence if (12) is satisfied,  (\ref{e20}) is satisfied as well. Although (\ref{e20}) is a better bound, we use (12) to make it independent of $\epsilon$

\subsection{Kullback Leibler Distance.}
\label{A:2}
The Kullback Leibler Distance (KD) is defined to be $-\sum_i{P(i)\log{\frac{Q(i)}{P(i)}}}$ where 
$$P(i)=\frac{x_{1i}-x_{2i}}{x_{4i}-x_{3i}}~~~ \mbox{and} Q(i)=\frac{r_{2i-1}(x_{1i}-x_{2i})+\epsilon_1x_{1i}-\epsilon_2x_{2i}}{r_{2i-1}(x_{4i}-x_{3i})+\epsilon_4x_{4i}-\epsilon_3x_{3i}}.$$
\begin{eqnarray}
KD &=& -\sum_i{\frac{x_{1i}-x_{2i}}{x_{4i}-x_{3i}}\log{\frac{r_{2i-1}+\frac{\epsilon_1x_{1i}-\epsilon_2x_{2i}}{x_{1i}-x_{2i}}}{r_{2i-1}+\frac{\epsilon_4x_{4i}-\epsilon_3x_{3i}}{x_{4i}-x_{3i}}}}} \nonumber \\
\label{e21}
 &=& \sum_i{\frac{x_{2i}-x_{1i}}{x_{4i}-x_{3i}}\log{\frac{r_{2i-1}+\frac{\epsilon_1x_{1i}-\epsilon_2x_{2i}}{x_{1i}-x_{2i}}}{r_{2i-1}+\frac{\epsilon_4x_{4i}-\epsilon_3x_{3i}}{x_{4i}-x_{3i}}}}}.
\end{eqnarray}
Without loss of generality, we assume that for $i=1$, the above expression attains minima,
$$\geq d\frac{x_{21}-x_{11}}{x_{41}-x_{31}}\log{\frac{r_{1}+\frac{\epsilon_1x_{11}-\epsilon_2x_{21}}{x_{11}-x_{21}}}{r_{1}+\frac{\epsilon_4x_{41}-\epsilon_3x_{31}}{x_{41}-x_{31}}}}$$
Let, $$D_1=\frac{KD}{ d\frac{x_{21}-x_{11}}{x_{41}-x_{31}}}\geq \log{\frac{r_{1}+\frac{\epsilon_1x_{11}-\epsilon_2x_{21}}{x_{11}-x_{21}}}{r_{1}+\frac{\epsilon_4x_{41}-\epsilon_3x_{31}}{x_{41}-x_{31}}}}.$$
Hence, $$e^{D_1} \geq  \frac{r_{1}+\frac{\epsilon_1x_{11}-\epsilon_2x_{21}}{x_{11}-x_{21}}}{r_{1}+\frac{\epsilon_4x_{41}-\epsilon_3x_{31}}{x_{41}-x_{31}}} \geq \frac{r_1-\frac{\epsilon x_{21}}{x_{11}-x_{21}}}{r_1+\frac{\epsilon x_{41}}{x_{41}-x_{31}}}.$$
Finally, $$KD \geq  d\frac{x_{11}-x_{21}}{x_{41}-x_{31}}\log{ \frac{r_1+\frac{\epsilon x_{41}}{x_{41}-x_{31}}}{r_1-\frac{\epsilon x_{21}}{x_{11}-x_{21}}}}.$$

\section{Range of bit length of the parameters.}
\label{B}
The probability of correctly guessing the random numbers from Equation (1) is computed as follows.
The adversary may arbitrarily fix the choice of two indices from $\{1,\ldots,2d\}$ for the $r_i$s
and the corresponding index from $\{1,\ldots,n\}$ for the choice of $\epsilon$.
Fixing the two $r_i$ from $2d$ many $r_i$'s can be done in ${2d\choose2}$ ways. Similarly choosing one
$\epsilon_i$ from $n$ many $\epsilon_i$'s can be done in $n$ ways. Hence the probability is:
$${2d \choose 2}{n \choose 1} \frac{1}{2^{2\ell_1}} \frac{1}{2^{\ell_2}}.$$ 
Similarly,the probability of correctly guessing from equation (13) is:
$${2d \choose 1}{n \choose 3}\frac{1}{2^{\ell_1}} \frac{1}{2^{3\ell_2}}.$$
Fixing $n$ and $d$ as chosen, for the probability to be less than $2^{-80}$, the following two equations must be satisfied,
\begin{equation}
\label{e24}
2\ell_1 +\ell_2 \geq 103,
\end{equation}
and
\begin{equation}
\label{e25}
\ell_1+3\ell_2 \geq 130.
\end{equation}
Hence the above two equations give us the range for the bit length of the parameters.

\end{document}